\documentclass[aps,prd,preprint,nofootinbib,superscriptaddress]{revtex4}
\usepackage{color}
\usepackage{graphicx}
\usepackage{psfrag}

\newcommand{\bq}{\mbox{\boldmath $q$}}
\newcommand{\bp}{\mbox{\boldmath $p$}}
\newcommand{\bk}{\mbox{\boldmath $k$}}
\newcommand{\br}{\mbox{\boldmath $r$}}
\newcommand{\bs}{\mbox{\boldmath $s$}}
\newcommand{\bt}{\mbox{\boldmath $t$}}
\newcommand{\bx}{\mbox{\boldmath $x$}}

\newcommand{\bP}{\mbox{\boldmath $P$}}
\newcommand{\bQ}{\mbox{\boldmath $Q$}}

\newcommand{\bR}{\mbox{\boldmath $R$}}

\newcommand{\brs}{\mbox{\scriptsize \boldmath $r$}}
\newcommand{\bss}{\mbox{\scriptsize \boldmath $s$}}
\newcommand{\bts}{\mbox{\scriptsize \boldmath $t$}}
\newcommand{\bxs}{\mbox{\scriptsize \boldmath $x$}}

\newcommand{\bqs}{\mbox{\scriptsize \boldmath $q$}}
\newcommand{\bks}{\mbox{\scriptsize \boldmath $k$}}

\newcommand{\bPs}{\mbox{\scriptsize\boldmath $P$}}
\newcommand{\bQs}{\mbox{\scriptsize\boldmath $Q$}}

\newcommand{\bRs}{\mbox{\scriptsize\boldmath $R$}}

\newcommand{\beqa}{\begin{eqnarray}}
\newcommand{\eeqa}{\end{eqnarray}}
\newcommand{\nn}{\nonumber}
\def\dfrac#1#2{\displaystyle\frac{#1}{#2}}
\begin{document}
\preprint{UTHEP-658}
\preprint{RIKEN-QHP-79}

\title{Asymptotic behavior of Nambu-Bethe-Salpeter wave functions for multi-particles in quantum field theories}

\newcommand{\TsukubaA}{
Graduate School of Pure and Applied Sciences, University of Tsukuba,
Tsukuba 305-8571, Japan
}

\newcommand{\TsukubaB}{
Center for Computational Sciences, University of Tsukuba, Tsukuba 305-8577, Japan
}

\newcommand{\TsukubaC}{
Kobe Branch, Center for Computational Sciences, University of Tsukuba, in RIKEN Advanced Institute for Computational Science(AICS), PortIsland, Kobe 650-0047, Japan
}

\newcommand{\Tokyo}{
Department of Physics, The University of Tokyo, Tokyo 113-0033, Japan
}

\newcommand{\Riken}{
Theoretical Research Division, Nishina Center, RIKEN, Wako 351-0198, Japan
}

\newcommand{\Nihon}{
Nihon University, College of Bioresource Sciences, Kanagawa 252-0880, Japan
}

\author{Sinya~Aoki\footnote{Address after April 1st, 2013: Yukawa Institute for Theoretical Physics, Kyoto University, Kitashirakawa Oiwakecho, Sakyo-ku, 
Kyoto 606-8502, Japan}}
\affiliation{\TsukubaA}
\affiliation{\TsukubaB}

\author{Noriyoshi~Ishii}
\affiliation{\TsukubaC}

\author{Takumi~Doi}
\affiliation{\Riken}

\author{Yoichi~Ikeda}
\affiliation{\Riken}

\author{Takashi~Inoue}
\affiliation{\Nihon}

%\author{(HAL QCD Collaboration)}

\begin{center}
\includegraphics[width=0.35\textwidth]{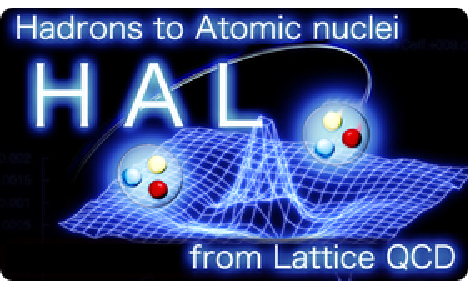}
\end{center}

\begin{abstract}
We derive asymptotic behaviors of the Nambu-Bethe-Salpeter (NBS) wave function at large space separations for systems with more than 2 particles in quantum field theories.
To deal with $n$-particles in the center of mass flame coherently, we introduce the Jacob coordinates of $n$ particles and then combine  their $3(n-1)$ coordinates into  the one spherical coordinate in $D=3(n-1)$ dimensions. We parametrize on-shell $T$-matrix for $n$-particle system of  scalar fields at low energy, using the unitarity constraint of the $S$-matrix.  
We then  express asymptotic behaviors of the NBS wave function for $n$ particles at low energy, in terms of
parameters of $T$-matrix, and show that the NBS wave function carry the information of $T$-matrix such as phase shifts and mixing angles of the $n$-particle system in its own asymptotic behavior, so that
the NBS wave function can be considered as  the scattering wave of $n$-particles in quantum mechanics. 
This property is one of the essential ingredients of the HAL QCD scheme to define "potential" from the NBS wave function in quantum field theories such as QCD.
Our results, together with an extension to systems with spin $1/2$ particles, justify the HAL QCD's definition of potentials for 3 or more nucleons(baryons) in terms the NBS wave functions.  
\end{abstract}

%\begin{document}
\maketitle

\section{Introduction}
\label{sec:introduction}
To understand hadronic interactions such as nuclear forces from the fundamental theory, Quantum Chromodynamics (QCD), non-perturbative methods such as the lattice QCD combined with numerical simulations 
are required, since the running coupling constant in QCD becomes large at hadronic scale.
Conventionally the finite size method\cite{Luscher:1990ux} has been employed to extract the scattering phase shift
in lattice QCD, but the method is so far limited to two-particle systems below the inelastic threshold,
except a few extensions\cite{Beane:2007es,Polejaeva:2012qv,Kreuzer:2012sr,Guo:2012hv}.

Recently an alternative method has been proposed  and  employed to extract the potential between nucleons  below inelastic thresholds\cite{Ishii:2006ec, Aoki:2008hh, Aoki:2009ji}.
This method has been extended, in order to investigate
other more general hadronic interactions such as baryon-baryon interactions\cite{Nemura:2008sp, Nemura:2009kc,Inoue:2010hs,Inoue:2010es,Inoue:2011ai} and meson-baryon interactions\cite{Ikeda:2010sg,Ikeda:2011qm,Kawanai:2010ev}.  See Refs.~\cite{Aoki:2011ep,Aoki:2012tk} for reviews of recent activities.

In the method, called the HAL QCD method, a potential between hadrons is defined in quantum field theories such as QCD, through the equal-time Nambu-Bethe-Salpeter (NBS) wave function\cite{Balog:2001wv} in the center of mass system,  which is defined for two nucleons as
\begin{eqnarray}
\Psi_{W}(\bx) &=& \langle 0 \vert T\left\{N(\br, 0) N(\br + \bx, 0) \right\}\vert NN,W\rangle_{\rm in}
\end{eqnarray}
where $\langle 0 \vert={}_{\rm out}\langle 0 \vert={}_{\rm in}\langle 0 \vert$ is the QCD vacuum (bra-)state,
$ \vert NN, W\rangle_{\rm in} $ is the two-nucleon asymptotic in-state at the total energy $W=2\sqrt{\bk^2+m_N^2}$ with the nucleon mass $m_N$ and the relative momentum $\bk$,  $T$ represents the time-ordered product, and $N(x)$ with $x=(\bx,t)$ is the nucleon operator. 
As the distance between two nucleon operators, $x= \vert \bx \vert$, becomes large, the NBS wave function satisfies the free Schr\"odingier equation, 
\begin{eqnarray}
\left( E_W - H_0 \right) \Psi_{W}(\bx) &\simeq& 0, \qquad  E_W=\frac{\bk^2}{2\mu},\quad H_0= \frac{-\nabla^2}{2\mu}
\label{eq:asymptotic}
\end{eqnarray}
where $\mu = m_N/2$ is the reduced mass. In addition, the asymptotic behavior of the NBS wave function is described in terms of the phase $\delta$ determined by the unitarity of the $S$-matrix,  $S=e^{2i\delta}$,
in QCD  (or the corresponding quantum field theory).  This has been shown for the elastic $\pi\pi$ scattering
\cite{Lin:2001ek, Aoki:2005uf}, where the partial wave of NBS wave function for the orbital angular momentum $L$ becomes
\begin{eqnarray}
\Psi_W^L &\simeq& A_L \frac{\sin ( k x - L\pi/2+\delta_L(W) )}{k x}
\label{eq:phase}
\end{eqnarray}
as $x\rightarrow \infty$ at $W\le W_{\rm th} = 4m_\pi$ (the lowest inelastic threshold).
The asymptotic behavior of the NBS wave function for the elastic $NN$ scattering has been derived in Ref.~\cite{Ishizuka2009a}.  

The HAL QCD method has also been applied to investigate three nucleon forces(3NF)\cite{Doi:2010yh,Doi:2011gq},  even though asymptotic behaviors of NBS wave function for three nucleons have not been derived yet.
The 3NF is necessary to explain  the experimental binding energies of 
light nuclei~\cite{Kamada:2001tv, Pieper:2007ax} and high precision deuteron-proton  elastic scattering data at intermediate energies~\cite{Sekiguchi:2011ku}.
It may also play an important role for various  phenomena
in nuclear physics and astrophysics~\cite{Furumoto:2009zz, Otsuka:2009cs,Akmal:1998cf, Nishizaki:2002ih}.

The purpose of this paper is to derive asymptotic behaviors of NBS wave functions for $n$ particles with $ n\ge 3$ at large distances where separations among $n$ operators become all large.
To avoid complications due to non-zero spins of particles, we consider  scalar fields in this paper. 
The results of this paper, together with an extension to spin $1/2$ particles, fills the  logical gap in the derivation 
of 3NF by the HAL QCD method\cite{Doi:2010yh,Doi:2011gq}.

In Sec.~\ref{sec:definition}, we explain our notations and definitions such as the modfied Jacobi coordinate, the Lippmann-Schwinger equation, and the NBS wave function for $n$ scalar particles.
In Sec.~\ref{sec:T-matrix} we  parametrize on-shell $T$-matrix for $n$ particles, by solving the unitarity constraint of $S$-matrix.
For explicit calculations for $n$-particle systems, we introduce the spherical coordinates in $D=3(n-1)$ dimensions, which is equal to a number of degrees of freedom for $n$ particle in 3-dimensions  in the center of mass flame, together with non-relativistic approximations.   
In Sec.~\ref{sec:NBS_asym},
using these techniques and results obtained in Sec.~\ref{sec:T-matrix},
we derive  asymptotic behaviors of NBS wave functions for $n$-particles, in terms of phase shifts and
mixing angles of the $n$-particle scattering. 
Conclusions and discussions  are given in Sec.~\ref{sec:conclusion}.
Some technical details are collected in three appendices.

\section{Some definitions and notations}
\label{sec:definition}
In this paper,  to avoid complications arising from nucleon spins, we consider an $n$-scalar particle system which have the same mass $m$  in the center of mass frame,
whose  coordinates and momenta are denoted by $\bx_i$, $\bp_i$ ($i=1,2,\cdots, n$ ) with $\displaystyle\sum_{i=1}^n \bp_i = 0$.
We introduce  modified Jacobi coordinates and corresponding momenta as
\beqa
\br_k=\sqrt{\frac{k}{k+1}}\times  \br_k^J, \qquad \bq_k =\sqrt{\frac{k+1}{k}}\times \bq_k^J
\eeqa
where the standard Jacobi coordinates and momenta are given by
\begin{eqnarray}
\br_k^J =\frac{1}{k}\sum_{i=1}^k {\bx}_i-{\bx}_{k+1}, \quad
\bq_k^J =\frac{k}{k+1}\left(\frac{1}{k}\sum_{i=1}^k {\bp}_i- {\bp}_{k+1} \right), 
\end{eqnarray}
for $k=1,2,\cdots, n-1$.  It is easy to see
\begin{eqnarray}
\sum_{i=1}^n {\bp}_i\cdot{\bx}_i &=&
\sum_{i=1}^{n-1} {\bq}_i\cdot {\br}_i , \quad
E=\frac{1}{2m}\sum_{i=1}^n {\bp}_i^2 =\frac{1}{2m}\sum_{i=1}^{n-1}{\bq}_i^2 .
\end{eqnarray}
The integration measure for modified Jacobi momenta is given by 
\begin{eqnarray}
\prod_{i=1}^n d^3p_i\, \delta^{(3)}\left(\sum_{i=1}^n {\bp}_i \right) =\frac{1}{n^{3/2}}\prod_{i=1}^{n-1} d^3 q_i .
\end{eqnarray}

\subsection{Lippmann-Schwinger equation}
\label{sec:LS}
As mentioned in the introduction, the asymptotic behavior of the NBS wave functions for a two-particle system has
already been derived in Refs.~\cite{Lin:2001ek,Aoki:2005uf,Ishizuka2009a,Aoki:2009ji}.
It is not straightforward, however,  to extend their derivations to multi-particle systems.
Instead, we utilize the Lippmann-Schwinger equation\cite{text_weinberg}, 
\begin{eqnarray}
\vert \alpha \rangle_{\rm in} &=& \vert \alpha\rangle_0 +
\int d\beta \frac{\vert \beta\rangle_ 0 T_{\beta\alpha}}{E_\alpha - E_\beta+i\varepsilon}, \qquad
T_{\beta\alpha} = {}_0\langle \beta \vert V \vert \alpha\rangle_{\rm in},
\label{eq:LS}
\end{eqnarray}
which is found to be a powerful tool  to study multi-particle systems.
We assume in this paper that no bound state appears in two or more particle systems.
Here the asymptotic in-state  $\vert \alpha\rangle_{\rm in}$ satisfies 
\begin{eqnarray}
(H_0 + V) \vert \alpha\rangle_{\rm in} &=& E_\alpha \vert \alpha\rangle_{\rm in},
\end{eqnarray}
whereas the non-interacting  state $\vert \alpha\rangle_0$ satisfies
\begin{eqnarray}
H_0  \vert \alpha\rangle_{0} &=& E_\alpha \vert \alpha\rangle_{0}.
\end{eqnarray}
The off-shell $T$-matrix  element or the "potential" $T_{\beta\alpha}= {}_0\langle \beta \vert V \vert \alpha\rangle_{\rm in}$ 
is related to the on-shell $S$-matrix element as
\begin{eqnarray}
S_{\beta\alpha}&\equiv&{}_{\rm out}\langle \beta \vert \alpha\rangle_{\rm in} \equiv
{}_0\langle \beta \vert S \vert \alpha\rangle_0
=\delta(\beta-\alpha) -2\pi i \delta(E_\alpha-E_\beta) T_{\beta\alpha}.
\end{eqnarray}
If we define $S= 1 -  i T$, we obtain
\begin{eqnarray}
{}_0\langle \beta \vert T \vert \alpha \rangle_0 &=& 2\pi  \delta(E_\alpha - E_\beta ) T_{\beta\alpha} .
\end{eqnarray}

%The relation between the two is given by\cite{text_weinberg}
%\begin{eqnarray}
%\int d\alpha\,  g(\alpha)\, e^{-iH t} \vert \alpha\rangle_{\rm in(out)}
%\rightarrow \int d\alpha\,  g(\alpha)\, e^{-iH_0 t} \vert \alpha\rangle_{0}
%\end{eqnarray}
%as $t\rightarrow -\infty(+\infty)$ for some (smearing function) $g(\alpha)$.

\subsection{NBS wave functions}
\label{sec:NBS}
The equal-time Nambu-Bethe-Salpeter(NBS) wave function for $n$ scalar particles is defined by
\begin{eqnarray}
\Psi^n_{\alpha}([{\bx}] ) &=& {}_{\rm in}\langle 0 \vert \varphi^n([ {\bx} ],0 ) \vert \alpha\rangle_{\rm in},
\end{eqnarray}
where
\begin{equation}
\varphi^n([ {\bx} ] , t) = T\{ \prod_{i=1}^n \varphi_i({\bx}_i,t) \},
\end{equation}
with the time-ordered product $T$, $[\bx ] =\bx_1,\bx_2,\cdots,\bx_n$, and $i$ represents a "flavor" of scalar field.
For simplicity, we regard all $n$ scalar particles are different but have the same mass $m$.

From the Lippmann-Schwinger equation (\ref{eq:LS}), the vacuum instate is given by
\begin{eqnarray}
\vert 0 \rangle_{\rm in} &=& \vert 0\rangle_0 + \int d\gamma \frac{ \vert \gamma \rangle_0 T_{\gamma 0}}{E_0-E_\gamma + i\varepsilon} .
\label{eq:vacuum}
\end{eqnarray}
As shown in Appemdix~\ref{app:vacuum}, however, the contribution from the second term to the
NBS wave function at large distances amounts to  
\begin{eqnarray}
{}_{\rm in}\langle 0 \vert \varphi^n([\bx],0)\vert \alpha \rangle _0 \simeq  \frac{1}{Z_\alpha} {}_0\langle 0 \vert \varphi^n([\bx],0)\vert \alpha \rangle _0 ,
\label{eq:plain_wave}
\end{eqnarray}
where $Z_\alpha$ is the normalization factor whose deviation from the unity comes from the off-shell $T$-matrix $T_{\gamma 0}$.
Using this and the Lippmann-Schwinger equation (\ref{eq:LS}),
the NBS wave function can be written as
\begin{eqnarray}
\Psi^n_{\alpha}([ {\bx}] ) &=&\frac{1}{Z_\alpha} {}_{0}\langle 0 \vert \varphi^n([ {\bx}],0  ) \vert \alpha\rangle_0 + \int d\beta \frac{1}{Z_\beta}\frac{{}_{0}\langle 0 \vert \varphi^n([ {\bx} ],0 ) \vert \beta\rangle_ 0 T_{\beta\alpha}}{E_\alpha - E_\beta+i\varepsilon} .
\label{eq:NBS}
\end{eqnarray}

To evaluate the above expression explicitly, we quantize all complex scalar fields in the Heisenberg representation at $t=0$ as
\begin{eqnarray}
\varphi_i({\bx},0) &=& \int\frac{d^3 k}{\sqrt{(2\pi)^3 2E_{k_i}}}\left\{a_i({\bk})e^{i{\bks\cdot \bxs}}+b_i^\dagger({\bk})e^{-i{\bks\cdot\bxs}}\right\} \\
\vert \alpha \rangle_0 &\equiv& \vert [{\bk}]_n\rangle_0 =\prod_{i=1}^n a_i^\dagger({\bk}_i) \vert 0\rangle_0,\quad
E_{k_i}=\sqrt{{\bf k}_i^2 + m^2},
\end{eqnarray}
where $[\bk ]_n =\bk_1,\bk_2,\cdots,\bk_n$ with $\sum_{i=1}^n\bk_i = 0$, and 
the full time evolution is given by $\varphi^n([\bx ],t) = e^{iH t}  \varphi^n([\bx],0) e^{-iH t}$ while
$H\rightarrow H_0$ for the free field. Our state normalization is given by
\begin{eqnarray}
{}_0\langle \beta_m \vert\alpha_n\rangle_0 &=&\delta(\beta_m -\alpha_n) .
\end{eqnarray}

Using the above, for the $n$ particle system in the center of mass frame, we have
\begin{eqnarray}
_0\langle 0\vert \varphi^n([{\bx}],0)\vert [\bk]_n \rangle_0 &=& \left(\frac{1}{\sqrt{(2\pi)^3}}\right)^n\prod_{i=1}^n
\frac{1}{\sqrt{2 E_{k_i}}} e^{i {\bks_i\cdot\bxs_i}}\nonumber \\
&=&
  \left(\frac{1}{\sqrt{(2\pi)^3}}\right)^n\left(\prod_{i=1}^n
\frac{1}{\sqrt{2 E_{k_i}}}\right) \exp\left[{i\displaystyle \sum_{j=1}^{n-1} {\bq}_j\cdot{\br}_j }\right],
\label{eq:plain}
\end{eqnarray}
where ${\br}_j$ and ${\bq}_j$ are modified Jacobi coordinates and momenta, respectively.

\section{Unitarity of $S$-matrix and parametrization of $T$-matrix}
\label{sec:T-matrix}
The unitarity of $S$-matrix implies
\beqa
T^\dagger - T = i T^\dagger T .
\eeqa
Defining
\beqa
_0\langle [{\bp^A}]_n \vert T \vert  [{\bp^B}]_n\rangle_0 &\equiv & 
\delta(E^A- E^B) \delta^{(3)}( {\bP^A} - {\bP^B}) T( [{\bq^A}]_n,  [{\bq^B}]_n )
\eeqa
where $[ \bp^X ]_n =\bp^X_1,\bp^X_2,\cdots, \bp^X_n$,  $[ \bq^X ]_n =\bq_1^X, \bq_2^X,\cdots, \bq_{n-1}^X $ with $X=A,B$, and 
\beqa
E^A \equiv\sum_{i=1}^n E_{p^A_i}, \quad E^B \equiv\sum_{i=1}^n E_{p^B_i}, \quad {\bP^A} \equiv\sum_{i=1}^n {\bp^A}_i,
\quad  {\bP^B} \equiv\sum_{i=1}^n {\bp^B}_i .
\eeqa
Here we parametrize the $T$-matrix element in terms of modified Jacobi momenta $[\bq^A ]$ and $[\bq^B]$.
Note that $T_{\beta\alpha}$, appeared in Lippmann-Schwinger equation, is expressed as
\beqa
T_{\beta\alpha} &=& \frac{1}{2\pi} \delta^{(3)}( {\bP^A} - {\bP^B}) T( [{\bq^A}]_n,  [{\bq^B}]_n ) .
\eeqa

Using the above expression, the unitarity constraint to $T$-matrix can be written as 
\beqa
T^\dagger( [{\bq^A}]_n,  [{\bq^B}]_n ) &-& T ( [{\bq^A}]_n,  [{\bq^B}]_n ) = \frac{i}{n^{3/2}}
\int \prod_{i=1}^{n-1}d^3q_i^C\, \delta(E^A-E^C) \nn \\
&\times& T^\dagger( [{\bq^A}]_n,  [{\bq^C}]_n )  T ( [{\bq^C}]_n,  [{\bq^B}]_n ) .
\label{eq:unitarity}
\eeqa
Our task is to solve this constraint.

\subsection{$n=2$}
Let me consider the simplest case, $n=2$.
In this case, we can parametrize $T$-matrix, in terms of the spherical harmonic functions $Y_{lm}$  as follows.
\beqa
T( {\bq^A},  {\bq^B} ) &=& \sum_{l,m} T_l(q^A,q^B) Y_{lm}(\Omega_{\bqs^A}) \overline{Y_{lm}(\Omega_{\bqs^B})}
\label{eq:T_expand}
\eeqa
where $q^{A,B}=\vert {\bq^{A,B}}\vert$ and $\Omega_{\bqs}$ is the solid angle of the vector ${\bq}$.
Using orthogonal property of $Y_{lm}$, the constraint becomes
\beqa
\overline{T_l}(q,q) - T_l(q,q) &=& \frac{i}{2^{3/2}}\int (q^C)^2 dq^C\, \delta(E-E_C) \overline{T_l}(q,q^C) T_l(q^C,q)
\eeqa
where $q= q^A=q^B$, $E=E^A=E^B=2\sqrt{m^2+q^2/2}$ and $E^C=2\sqrt{m^2+(q^C)^2/2}$.
After $q^C$ integral, the constraint  now becomes
\beqa
\overline{T_l}(q,q) - T_l(q,q) &=& i\frac{q E}{2\times 2^{3/2}}  \overline{T_l}(q,q) T_l(q,q),
\eeqa
which can be solved as
\beqa
T_l(q)\equiv T_l(q,q) = -\frac{4\times 2^{3/2}}{q E} e^{i\delta_l(E)}\sin \delta_l(E),
\label{eq:T2}
\eeqa
where $\delta_l(q)$ is the phase shift for the partial wave with the angular momentum $l$ at energy $E
=2\sqrt{m^2+q^2/2}$. 

\subsection{General $n$}
For general $n$ case, we introduce the non-relativistic approximation for the energy in the delta-function as
\beqa
E^A-E^C &\simeq& \frac{(p^A)^2-(p^C)^2}{2m} =\frac{(q^A)^2-(q^C)^2}{2m} 
\eeqa
where $(q^{A,C})^2=\sum_{i=1}^{n-1} (\bq_i^{A,C})^2$ for modified Jacobi momenta $[\bq^{A,C}]_n$.
To perform 3 dimensional momentum integral $(n-1)$ times, we consider $D=3(n-1)$ dimensional space.
Denoting $s=\vert {\bs}\vert$ is a $D$-dimensional hyper-radius and $\Omega_{\bss}$ are angular variables for the vector ${\bs}$ in $D$ dimensions, the Laplacian operator is decomposed as
\beqa
\nabla^2 &=& \frac{\partial^2}{\partial s^2} +\frac{D-1}{s}\frac{\partial}{\partial s} - \frac{\hat L^2}{s^2}
\eeqa
where $\hat L^2$ is angular-momentum  in $D$-dimensions.
The hyper-spherical harmonic function\cite{Avery1993}, an extension of spherical harmonic function in 3-dimension to general $D$-dimensions satisfies
\beqa
\hat L^2 Y_{[L]}(\Omega_{\bss}) &=& L(L+D-2) Y_{[L]}(\Omega_{\bss})
\eeqa
where $[L]=L,M_1,M_2,\cdots$ are a set of "quantum" numbers specifying  the hyper-spherical harmonic function. The hyper-spherical harmonic function is orthogonal such that
\beqa
\int d\Omega_{\bss}\, \overline{Y_{[L]}(\Omega_{\bss})} Y_{[L^\prime]}(\Omega_{\bss}) &=& \delta_{[L][L^\prime]}
\eeqa
and complete
\beqa
\sum_{[L]}\overline{Y_{[L]}(\Omega_{\bss})} Y_{[L]}(\Omega_{\bts})\delta(s-t) &=& s^{D-1}\delta^{(D)}({\bs}-{\bt}),
\eeqa
so that an arbitrary function $f({\bf  s})$ of ${\bf s} \in R^D$ can be expanded as
\beqa
f({\bs}) &=& \sum_{[L]} f_{[L]} (s) Y_{[L]}(\Omega_{\bss}) .
\eeqa

Using the hyper spherical function, we expand the $T$-matrix as
\beqa
T([{\bq^A}]_n, [{\bq^B}]_n) &\equiv& T({\bQ_A}, {\bQ_B}) \nn \\
&=&\sum_{[L],[K]} T_{[L][K]}(Q_A,Q_B) Y_{[L]}(\Omega_{\bQs_A})\overline{Y_{[K]}(\Omega_{\bQs_B})}
\label{eq:T_expand_n}
\eeqa
where ${\bQ_X} =({\bq^X}_1, {\bq^X}_2, \cdots , {\bq^X}_{n-1} )$ for $X=A,B$ is a momentum vector in $D=3(n-1)$ dimensions. \footnote{For $n\ge 3$, the $T$-matrix  can have singularities at particular on-shell values of external momenta\cite{Potapov:1977ux,Potapov:1977sr,Dashen:1971ab}, which are expressed in terms of delta functions and principles values with the $i\varepsilon$ prescription for propagators.
Even for such cases,  however,  
our expansion of $T$-matrix in eq.~(\ref{eq:T_expand_n}) is still valid in the sense of distributions,
and these singularities originate from a sum over infinite terms\cite{Newton:1974ec}. 
We would like to thank Prof. S.~R.~Sharpe and Dr. M.~T.~Hansen for pointing out this problem and relevant references.
 }

With the non-relativistic approximation and orthogonal property, 
the unitarity relation eq.~(\ref{eq:unitarity}) after  $\Omega_{\bQs^C}$ integration leads to 
\beqa	
T^\dagger_{[L][K]}(Q_A, Q_A) - T_{[L][K]}(Q_A, Q_A) &=&
\frac{i}{n^{3/2}}\int Q^{D-1} dQ\,  \delta(E_A-E)\, 
T^\dagger_{[L][N]}(Q_A, Q) T_{[N][K]}(Q, Q_A) \nn \\
&=& i \frac{  m (Q_A)^{D-2}}{n^{3/2}} \sum_{[N]}T^\dagger_{[L][N]}(Q_A, Q_A) T_{[N][K]}(Q_A, Q_A) 
\eeqa
where $Q_A=Q_B$ is used. 
By diagonalizing $T$ with an unitary matrix $U$ as
\beqa
T_{[L][K]}(Q,Q) &=& \sum_{[N]} U_{[L][N]}(Q) T_{[N]}(Q) U_{[N][K]}^\dagger(Q),
\label{eq:UTU}
\eeqa
the above constraint can be solved as
\beqa
T_{[L]}(Q) = - \frac{2 n^{3/2}}{m Q^{3n-5}} e^{i\delta_{[L]}(Q)} \sin \delta_{[L]}(Q), 
\label{eq:Tn}
\eeqa
where $\delta_{[L]} (Q)$ is a real phase, which depends on $Q$ and $[L]$ in $D=3(n-1)$ dimensions.  
This is a main result of this section.
Unfortunately, a relation of the phase shifts in the hyper-spherical coordinates with physical observables for $n$-particles  in the standard Jacobi coordinates is not transparent.
Therefore it will be an important task in the future  to make the relation between them clear. 

At $n=2$, we have
\beqa
T_{[L]}(Q) = -\frac{2\times 2^{3/2}}{m Q} e^{i\delta_{[L]}(Q)}\sin \delta_{[L]}(Q), 
\eeqa
which agrees with eq.~(\ref{eq:T2})  under the non-relativistic approximation that $E\simeq 2 m$,
together with the replacement that $Q\rightarrow q$ and $[L]\rightarrow l$ and $U\rightarrow 1$.

\section{Asymptotic behaviors of NBS wave functions for $n$ particles}
\label{sec:NBS_asym}
In this section, we derive the asymptotic behaviors of NBS wave functions for multi-particle systems, using 
expressions~(\ref{eq:NBS})
\beqa
\Psi^n_{\alpha}( [\bx ] ) &=&\frac{1}{Z_\alpha} {}_{0}\langle 0 \vert \varphi^n([\bx],0  ) \vert \alpha\rangle_0 + \int d\beta \frac{1}{Z_\beta}\frac{{}_{0}\langle 0 \vert \varphi^n([ {\bx} ],0 ) \vert \beta\rangle_ 0 T_{\beta\alpha}}{E_\alpha - E_\beta+i\varepsilon} 
\eeqa  
and (\ref{eq:plain})
\beqa
_0\langle 0\vert \varphi^n([{\bx}],0)\vert [\bk]_n \rangle_0 
&=&
  \left(\frac{1}{\sqrt{(2\pi)^3}}\right)^n\left(\prod_{i=1}^n
\frac{1}{\sqrt{2 E_{k_i}}}\right) \exp\left[{i\displaystyle \sum_{j=1}^{n-1} {\bq}_j\cdot{\br}_j }\right] .
\eeqa

\subsection{$n=2$}
As an exercise, let us first consider the $n=2$ case, whose result is already known.
Using $\br =(\bx_2-\bx_1)/\sqrt{2}$, $\bp_1=-\bp_2 = \bq/\sqrt{2}$ and $E_q=\sqrt{m^2+q^2/2}$, the NBS wave function at $n=2$ is given by
\beqa
\Psi^2_{\bqs}(\br ) &=& \frac{1}{ 2 E_q Z_q}\left[ \frac{e^{i\bqs\cdot\brs} }{(2\pi)^3}
+ \int \frac{d^3 k}{2^{3/2} (2\pi)^3}\frac{Z_q E_q}{Z_k E_k}\frac{e^{i\bks\cdot\brs}\,T(\bk,\bq)}{4\pi (E_q-E_k+i\varepsilon)}
\right] ,
\label{eq:2particle1}
\eeqa
where $\bk$ is also the modified Jacobi momentum.
Using expansions that
\beqa
e^{i\bqs\cdot\brs} &=& 4\pi \sum_{lm} i^l j_l(qr) Y_{lm}(\Omega_{\brs})\, \overline{Y_{lm}(\Omega_{\bqs})} 
\label{eq:plain_3}\\
\Psi^2_{\bqs}(\br ) &=&  \sum_{lm} i^l \Psi^2_l(r,q) Y_{lm}(\Omega_{\brs})\, \overline{Y_{lm}(\Omega_{\bqs})}, 
\eeqa
where $j_l(x)$ is the spherical Bessel function of the first kind, 
together with eq.~(\ref{eq:T_expand}), and integrating  over $\Omega_{\bks}$, we obtain
\beqa
 \Psi^2_l(r,q)&=&  \frac{4\pi}{ (2\pi)^3 2 E_q Z_q}\left[ j_l(qr) +\int_0^\infty \frac{k^2dk}{2\pi 2^{3/2}} \, \frac{Z_q E_q}{Z_k E_k}
 \frac{j_l(kr)T_l(k,q)}{2(E_q-E_k+i\varepsilon)}\right] .
 \label{eq:NBS_2}
\eeqa
Since $E_q$ is below inelastic thresholds, we assume that $T_l(q,k)$ does not have any poles in the positive real axis. Under this assumption, we perform the $k$ integral using the formula
\beqa
\int_{0}^{\infty} k^2 dk  \frac{ j_l(kr)}{q^2-k^2+i\varepsilon} F_l(k) &\simeq& -\frac{\pi q^2}{2q} F_l(q) [ n_l(qr)+i j_l(qr) ]
\label{eq:Bessel_int}
\eeqa
for $r\gg 1$\cite{Aoki:2009ji,Aoki:2005uf}, where $F_l(k)$ does not have any poles in the positive real axis and satisfies $\displaystyle \int k^{-l} j_0(kr) F_l(k)  k^2 d k  \simeq 0$, which follows from  $(\nabla^2+q^2)\Psi^2_{\bqs}(\br) \simeq 0$, for large $r$\cite{Aoki:2009ji}, 
 and $n_l(x)$ is the spherical Bessel function of the second kind.
After the $k$ integral using this formula,
the second term  in eq.~(\ref{eq:NBS_2})  becomes
\beqa
-\left[n_l(qr) + i j_l(qr) \right] \frac{q E_q}{2\times 2^{3/2}} T_l(q,q)
&=& \left[n_l(qr) + i j_l(qr) \right] e^{i\delta_l(q)}\sin \delta_l(q) ,
\eeqa
where  
the unitarity constraint (\ref{eq:T2}) for $T_l(q,q)$ is used to obtain the last equality. 
We then obtain
\beqa
 \Psi^2_l(r,q)&=&  \frac{4\pi}{ (2\pi)^3 2 E_q Z_q}e^{i\delta_l(q)}\left[ j_l(qr) \cos\delta_l(q) + n_l(qr)\sin\delta_l(q)\right] \\
 &\simeq &  \frac{4\pi}{ (2\pi)^3 2 E_q Z_q}\frac{e^{i\delta_l(q)}}{qr} \sin ( qr -l\pi/2 +\delta_l(q) ) 
 \label{eq:asym_2}
\eeqa
for $r\gg 1$, where asymptotic behaviors that $j_l(x) \simeq \sin(x-l\pi/2)/x$ and $n_l(x) \simeq \cos(x-l\pi/2)/x$ are used.
The phase of the $S$-matrix,  $\delta_l(q)$,  can be interpreted as the scattering phase shift of the NBS wave function for the $n=2$ case.

\subsection{General $n$}
The NBS wave function for general $n$ is expressed as
\beqa
\Psi^n(\bR, \bQ_A ) &=& C(\bQ_A) \left[ e^{i\bQs_A\cdot \bRs} + \frac{n^{-3/2}}{2\pi} \int d^D\, Q \frac{C(\bQ)}{C(\bQ_A)}
\frac{e^{i\bQs\cdot \bRs}}{E_{Q_A} - E_Q +i\varepsilon} T(\bQ, \bQ_A)\right]
\eeqa
where $\bR=(\br_1,\br_2,\cdots, \br_{n-1})$ and $\bQ^{(A)} =(\bq_1,\bq_2,\cdots, \bq_{n-1})^{(A)}$ are the modified Jacobi coordinates and momenta in $D=3(n-1)$ dimensions,
\beqa
C(\bQ_A) &=& \frac{1}{Z(\bQ_A)}\prod_{j=1}^n \frac{1}{\sqrt{(2\pi)^32E_{p^A_j}}} ,
\eeqa
with $\bp_j^{(A)}$ is the momentum of the $j$-th particle.  In the non-relaivistic limit that
\beqa
C(\bQ_A) &\rightarrow& C(Q_A)= \frac{1 +c \dfrac{Q_A^2}{m}}{( (2\pi)^3 2m )^{n/2}}
, \quad \frac{C(\bQ)}{C(\bQ_A)} \rightarrow \frac{C(Q)}{C(Q_A)}, \quad
(E_{Q_A} - E_Q) \rightarrow \frac{Q_A^2-Q^2}{2m}
\eeqa
with some constant $c$,
we have
\beqa
\Psi^n(\bR, \bQ_A ) &=& C(Q_A)\left[ e^{i\bQs_A\cdot \bRs} + \frac{2 m}{2\pi n^{3/2}}\int d^D Q\,
\frac{C(Q)}{C(Q_A)}\frac{  e^{i\bQs\cdot \bRs}}{ Q_A^2-Q^2 +i\varepsilon} T(\bQ,\bQ_A)\right] .
\label{eq:NBS_nonR}
\eeqa

In $D$-dimensions, we have\cite{Avery1993}
\beqa
e^{i\bQs\cdot\bRs} &=& (D-2)!! \frac{2\pi^{D/2}}{\Gamma(D/2)}\, \sum_{[L]}\, i^L\, j_L^D(QR)\, Y_{[L]}(\Omega_{\bRs})\,\overline{Y_{[L]}(\Omega_{\bQs})},
\label{eq:plain_D}
\eeqa
which is the generalization of the $D=3$ formula in eq.~(\ref{eq:plain_3}),
where $j_L^D$ is the hyperspherical Bessel function of the first kind defined by
\beqa
j_L^D(x) &=& \frac{\Gamma(\frac{D-2}{2})\ 2^{\frac{D-4}{2}}}{(D-4)!! \ x^{\frac{D-2}{2}}}\, J_{L_D}(x),
\eeqa
with $L_D = L +\frac{D-2}{2}$ and  the Bessel function of the first kind, $J_{L_D}(x)$.

Using an expansion that
\beqa
\Psi^n(\bR,\bQ_A) &=& \sum_{[L],[K]} \Psi^n_{[L],[K]}(R, Q_A) Y_{[L]}(\Omega_{\bRs})\overline{Y_{[K]}(\Omega_{\bQs_A})},
\eeqa
with eqs.~(\ref{eq:T_expand_n}) and (\ref{eq:plain_D}), 
and performing $d\, \Omega_{\bQs}$ integral, we obtain
 \beqa
  \Psi^n_{[L],[K]}(R, Q_A) &=& C(Q_A)  \frac{i^L(2\pi)^{D/2}}{ (Q_A R)^{\frac{D-2}{2}}}
  \left[ J_{L_D}(Q_A R) \delta_{LK} 
  +  \int d Q\, \frac{ J_{L_D} (QR) }{Q_A^2 - Q^2 +i\varepsilon} H_{[L],[K]}(Q, Q_A) \right] \nn \\
 \label{eq:Q-integral} 
 \eeqa
 where
 \beqa
 H_{[L],[K]}(Q, Q_A) &=& \frac{m}{\pi n^{3/2}} \frac{C(Q)}{C(Q_A)} Q ^{D/2} Q_A^{D/2-1} T_{[L],[K]}(Q, Q_A) .
 \label{eq:defH}
 \eeqa
 We now perform the $Q$ integral, assuming that  $T_{[L],[K]}(Q, Q_A)$ does not have any poles on the positive real axis at $Q_A$ below inelastic thresholds. We consider $n=2k$ and $n=2k+1$ cases separately.
 
 \subsubsection{$n=2k$ case}
 In this case, 
 \beqa
 J_{L_D}(x) = j_{L_k}(x)\,  \sqrt{\frac{2}{\pi}}\, x^{1/2}
 \eeqa
 where $L_k=L+3(k-1)$ and $j_{L_k}$ is the spherical Bessel function of the first kind.
Using eq.~(\ref{eq:Bessel_int}), the second term in eq.~(\ref{eq:Q-integral}) can be evaluated as\cite{Aoki:2009ji,Aoki:2005uf}
 \beqa
 &&\int dQ\,  \frac{ j_{L_k} (QR) }{Q_A^2 - Q^2 +i\varepsilon} \sqrt{\frac{2}{\pi}}\, (QR)^{1/2} H_{[L],[K]}(Q, Q_A) \nn \\
& \simeq&  - \left[ n_{L_k}(Q_A R) + i j_{L_k}(Q_A R)\right]  \frac{\pi}{2Q_A} \sqrt{\frac{2}{\pi}}\, (Q_AR)^{1/2} H_{[L],[K]}(Q_A, Q_A) \nn \\
& =&
 \left[ N_{L_D}(Q_A R) + i J_{L_D}(Q_A R)\right] \sum_{[N]}U_{[L][N]}(Q_A) e^{i \delta_{[N]}(Q_A)}\sin \delta_{[N]}(Q_A) U_{[N][K]}^\dagger(Q_A)  
\eeqa
for $R\gg 1$, 
where the unitarity constraint to $T$ in eq.~(\ref{eq:Tn}) is used to obtain the last line, and 
$J_{L_D}$ and $N_{L_D}$ are Bessel functions of the first and second kinds, respectively.

\subsubsection{$n=2k+1$ case}
In this case, $L_D = L+3k-1$ is an integer, and for large $R$, $J_{L_D}(x)$ becomes
\beqa
J_{L_D} (x) &\simeq &\sqrt{\frac{2}{\pi  x}}\sin \left( x -\Delta_L\right), \
N_{L_D} (x) \simeq \sqrt{\frac{2}{\pi  x}}\cos \left( x -\Delta_L\right), \quad
\Delta_L = \frac{2L_D-1}{4}\pi .
\label{eq:bessel_asym}
\eeqa
Using this asymptotic behavior, the $Q$ integral in eq.~(\ref{eq:Q-integral})  can be performed, and we obtain for $R\gg 1$ 
\beqa
I&\equiv &\int dQ\, \frac{J_{L_D}(QR)}{Q_A^2-Q^2+i\varepsilon}H_{[L],[K]}(Q,Q_A) \nn \\
&\simeq& -\sqrt{\frac{2}{\pi  Q_A R}}\left[ \frac{\pi e^{i (Q_A R -\Delta_L)}}{2 Q_A} H_{[L],[K]}(Q_A,Q_A)+O\left(R^{(3-D)/2}\right)\right]
\label{eq:integral} \\
&\simeq & \left[ N_{L_D}(Q_A R) + i J_{L_D}(Q_A R)\right] \sum_{[N]}
U_{[L][N]}(Q_A) e^{i\delta_{[N]}(Q_A)}\sin \delta_{[N]}(Q_A) U_{[N][K]}^\dagger(Q_A),
\eeqa
where, in the last line,  the $O(1/R)$ contribution is neglected for large $R$  and the unitarity condition for $T$ in eq.~(\ref{eq:Tn}) is used, and $e^{i (Q_A R -\Delta_D)}$ is replaced by the asymptotic behaviors of $J_n$ and $H_n$. The detailed calculation of the $Q$ integral is given in Appendix ~\ref{app:Q-int}.

\subsection{Asymptotic behavior}
For both $n=2k$ and $n=2k+1$, 
we finally obtain
\beqa
 \Psi^n_{[L],[K]}(R, Q_A) &\simeq& C i^L \frac{(2\pi)^{D/2}}{ (Q_A R)^{\frac{D-2}{2}}}\sum_{[N]}
 U_{[L][N]}(Q_A) e^{i \delta_{[N]}(Q_A)} U_{[N][K]}^\dagger(Q_A)   
\nn \\
 &\times&
 \left[
 J_{L_D}(Q_AR) \cos \delta_{[N]}(Q_A) + N_{L_D}(Q_A R) \sin \delta_{[N]}(Q_A) 
 \right] \\
 &\simeq&
 C i^L \frac{(2\pi)^{D/2}}{ (Q_A R)^{\frac{D-1}{2}}}\sum_{[N]}
 U_{[L][N]}(Q_A) e^{i \delta_{[N]}(Q_A)} U_{[N][K]}^\dagger(Q_A) \nn \\
&\times&  \sqrt{\frac{2}{\pi  }}\ \sin\left(Q_A R -\Delta_L +\delta_{[N]}(Q_A)\right)
 \label{eq:main_result}
\eeqa
for $R\gg 1$, which agrees with eq.~(\ref{eq:asym_2}) at $n=2$.
Eq.~(\ref{eq:main_result}) is the main result of this paper, which
tells us that the NBS wave function of $n$-particles for large $R$ can be considered as the generalized scattering wave of $n$ particles, whose generalized scattering phase shift $\delta_{[N]}(Q_A)$ is nothing but the phase of the $S$-matrix in QCD, determined  in eq.~(\ref{eq:Tn}) by the unitarity.

\section{Conclusion and discussion}
\label{sec:conclusion}
In this paper,  we have investigated the asymptotic behaviors of the NBS wave functions at large separations for $n$ complex scalar fields. 
We have first solved  the unitarity constraint of the S-matrix for $n\ge 3$, using the $D=3(n-1)$ coordinate space and employing the hyper-spherical harmonic function, together with the non-relativistic approximation for the energy. The results are summarized in eqs.~(\ref{eq:UTU}) and (\ref{eq:Tn}). We then have calculated the asymptotic behaviors of the NBS wave functions at large separations for $n\ge 3$,  using again the hyper-spherical harmonic function, which is found to be quite useful for this purpose.
We finally obtain eq.~(\ref{eq:main_result}), 
which is the main result in this paper.
In appendix~\ref{sec:coupled}, we generalize our results to the coupled channels, where the particle mixing occurs during the scattering.   

Using the results in this paper, we can generalize the HAL QCD method to hadron interactions for the $n$-particle system with $n \ge 3$.
This give a firm theoretical background to the extraction of interactions among many hadrons by the HAL QCD method, in particular,  the three nucleon force\cite{Doi:2010yh,Doi:2011gq},
together with an extension to systems with spin $1/2$ particles, which is a straightforward but much more complicated task in future.
Moreover, combining it with the results in our previous paper\cite{Aoki:2012bb},
which shows that non-local but energy independent potentials can be constructed from the NBS wave functions above the inelastic threshold,
the HAL QCD method can be extended to hadronic interactions above the inelastic threshold energy, where  particle productions such as $NN\rightarrow NN\pi$ can occur.

\section*{Acknowledgement}
S.A. would like to thank the Galileo Galilei Institute for Theoretical Physics for its kind hospitality during completion of this paper while attending the workshop "New Frontiers Lattice Gauge Theory" .
This work is supported in part by the Grants-in-Aid for Scientific Research (No. 24740146), the Grant-in-Aid for Scientific Research on Innovative Areas (No. 2004: 20105001, 20105003) and SPIRE (Strategic Program for Innovative Research).

\appendix
\section{Contribution from vacuum}
\label{app:vacuum}
In this appendix, we show eq.~(\ref{eq:plain_wave}). 
Assuming each flavor is conserved, ${}_0\langle  \gamma \vert $ which contributes in  eq.~(\ref{eq:plain_wave}) is
a sum of the following form.
\beqa
{}_0\langle  I_k \vert &=& 
 {}_0 \langle  0 \vert \prod_{i\in I_k}  a_{i}(\bk_i^A) b_{i}(\bk_i^B)
\eeqa
with $\sum_{i\in I_k} (\bk_i^A+\bk_i^B)=\boldmath{0}$,  where $ k \le n$ and $I_k=\{i_1,i_2,\cdots, i_k\}$ with $1\le i_1 < i_2 < \cdots < i_k \le n$. Note that the operator $a_i b_i$ creates a particle-antiparticle pair with flavor $i$.
Using this notation, we have
\beqa
{}_{\rm in} \langle 0 \vert \varphi^n([\bx],0)\vert [\bk]_n \rangle_0 &=&
{}_{0} \langle 0 \vert \varphi^n([\bx],0)\vert  [\bk]_n\rangle_0 
+\sum_{k=1}^n \sum_{I_k} \prod_{i\in I_k} \int d^3k_i^A d^3 k_i^B \nn \\
&\times& \delta^{(3)} \left(\sum_{i\in I_k} (\bk_i^A+\bk_i^B) \right)
 \frac{T^\dagger_{0 I_k}}{E_0-E_{I_k}+i\varepsilon}\ {}_0\langle  I_k \vert \varphi^n([\bx],0)\vert [\bk]_n \rangle_0,
\label{eq:plain_wave2}
\eeqa
where $E_0=0$ for the vacuum.

Using 
\beqa
(2\pi)^{3n/2} {}_0\langle  I_k \vert \varphi^n([\bx],0)\vert [\bk]_n \rangle_0 &=&
\prod_{i\in I_k}  \frac{e^{-i\bks_i^B \bxs_{i}} }{\sqrt{2E_{k_i^B}}} \delta^{(3)}(\bk_i-\bk_i^A)
%\nn \\
%&\times& 
%\delta^{(3)}\left(\sum_{i\in I_k}  (\bk_{i}^A + \bk_i^B) \right)
  \prod_{j\in \bar I_k} \frac{e^{i\bks_j \bxs_j}}{\sqrt{2E_{k_j}}}, 
\eeqa
where $\bar I_k\cup I_k = \{1,2,3,\cdots, n \} $ and $\bar I_k \cap I_k =\phi $, the second term in eq.~(\ref{eq:plain_wave2}) becomes
\beqa
C_n \sum_{k=1}^n \sum_{I_k} \prod_{i\in I_k}
\int \frac{d^3 k_i^B} {\sqrt{2E_{k_i^B}}} e^{-i\bks_i^B \bxs_{i}} \delta^{(3)}\left(\sum_{i\in I_k}  (\bk_{i} + \bk_i^B) \right)
  \prod_{j\in \bar I_k} \frac{e^{i\bks_j \bxs_j}}{\sqrt{2E_{k_j}}}
\times  \frac{ T^\dagger_{0; I_k}(\boldmath{0};[\bk, \bk^B] )}{E_0 - E_{ [\bks, \bks^B]}},
\label{eq:second}
\eeqa
where $C_n=(2\pi)^{ -\frac{3n}{2}}$, $E_{ [\bks, \bks^B]} =\displaystyle \sum_{i\in I_k}\left( \sqrt{\bk_i^2+m^2} + \sqrt{(\bk^B_i)^2+m^2} \right)$,  $T^\dagger_{0; I_k}(0; [\bk, \bk^B] )$ is the off-shell T-matrix from vacuum to $2k$ particles,  and $[\bk, \bk^B] = \{ \bk_{i_1}, \bk^B_{i_1}, \bk_{i_2}, \bk^B_{i_2},\cdots, \bk_{i_k}, \bk^B_{i_k} \}$.

We first show that terms at $k \ge 2$ in eq.~(\ref{eq:second}) do not contribute at large distances.
After $\bk_{i_k}^B$ integral, the factor in the first exponential is written as
$\displaystyle
-i\sum_{i\in I_{k-1} }\bk_{i}^B (\bx_i - \bx_{i_k}) + i\sum_{i\in I_k} \bk_i \bx_{i_k}
$, where $I_{k-1} = \{i_1,i_2,\cdots, i_{k-1} \}$. Since $I_{k-1} \not= \phi $ for $k\ge 2$, we pefrom the $\bk_{i_1}^B$   integral in eq.~(\ref{eq:second}). Using the same method which leads to eq.~(\ref{eq:Bessel_int}) from eq.~(\ref{eq:2particle1}) and noticing the fact that there is no real poles for the $\bk_{i_1}^B$ integral in eq.~(\ref{eq:second}), it is clear that the contribution is suppressed exponentially in large $\vert \bx_{i_1}-\bx_{i_k} \vert$. This means that terms at $k=1$ only contribute in  eq.~(\ref{eq:second}) and other terms at $k\ge 2$ are suppressed asymptotically at large distances.  

The term at $k=1$ is easily evaluated as
\beqa 
C_n \prod_{j=1}^n  \frac{e^{i\bks_j \bxs_j}}{\sqrt{2E_{k_j}}} \sum_{i=1}^n
\frac{T_{0;i-i}^{\dagger}( 0; \bk_i, -\bk_i )}{ -2\sqrt{\bk_i^2+m^2}} ,
\eeqa
where $T_{0;i,-i}$ is the off-shell T-matrix from vacuum to a pair of particle-anttiparticle with the flavor $i$.

We then finally obtain
\beqa
{}_{\rm in} \langle 0 \vert \varphi^n([\bx],0)\vert [\bk]_n \rangle_0 
&\simeq& \frac{1}{Z([\bk]_n)} {}_{0} \langle 0 \vert \varphi^n([\bx],0)\vert  [\bk]_n\rangle_0 
\eeqa 
with
\beqa
 \frac{1}{Z([\bk]_n)} &=& 1 + \sum_{i=1}^n
\frac{T_{0;i-i}^{\dagger}( 0; \bk_i, -\bk_i )}{- 2\sqrt{\bk_i^2+m^2}} ,
\eeqa
which proves eq.~(\ref{eq:plain_wave}) with $Z_\alpha = Z([\bk]_n)$.

\section{$Q$ integrals}
\label{app:Q-int}
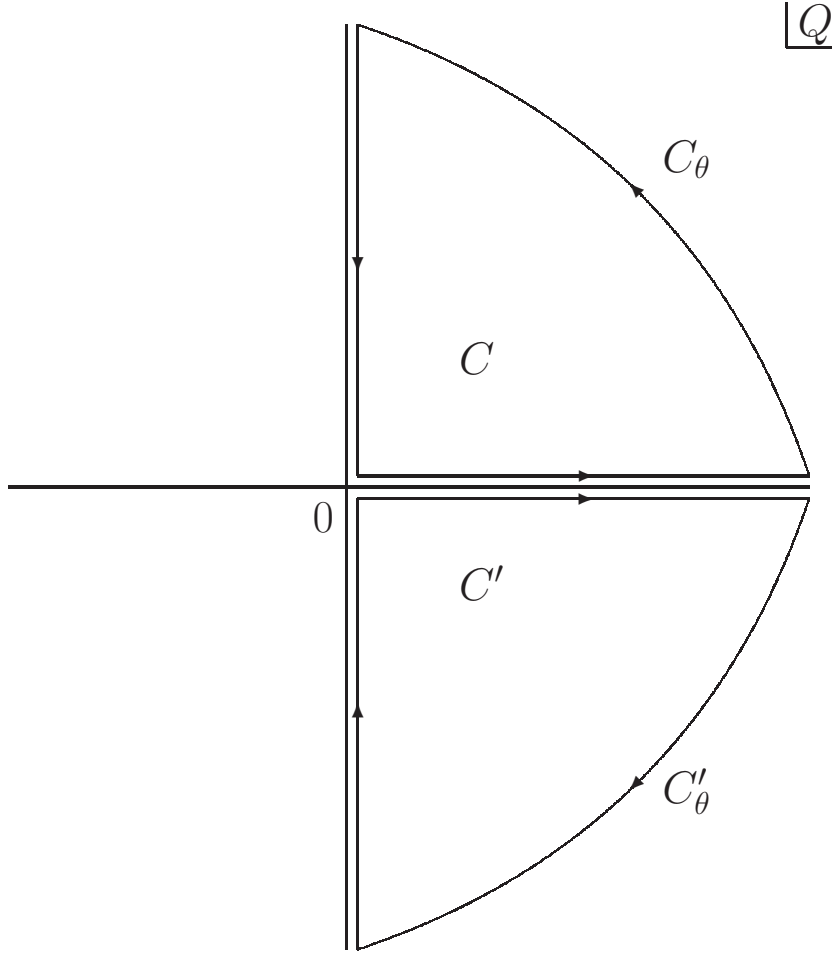
\begin{figure}[htb]
\begin{center}
\setlength{\unitlength}{1.5mm}
\begin{picture}(82,82)(0,-1)
\put(0,40){\thicklines\line(1,0){71}}
\put(30, -1){\thicklines\line(0,1){82}}
\put(27,36){\Large $0$}
\put(70,80){\Large$Q$}
\put(69,79){\thicklines\line(1,0){4}}
\put(69,79){\thicklines\line(0,1){4}}
\put(31,41){\thicklines\line(1,0){40}}
\put(50,41){\thicklines\vector(1,0){2}}
\put(31,41){\thicklines\line(0,1){40}}
\put(31,61){\thicklines\vector(0,-1){2}}
\put(40,50){\Large$C$}
\put(58,68){\Large$C_\theta$}
\bezier{600}(31,81)(61,71)(71,41)
\put(56,66){\thicklines\vector(-1,1){1}}
\put(31,39){\thicklines\line(1,0){40}}
\put(50,39){\thicklines\vector(1,0){2}}
\put(31,39){\thicklines\line(0,-1){40}}
\put(31,19){\thicklines\vector(0,1){2}}
\put(40,30){\Large$C^\prime$}
\put(58,12){\Large$C^\prime_\theta$}
\bezier{600}(31,-1)(61,9)(71,39)
\put(56,14){\thicklines\vector(-1,-1){1}}
\end{picture}
\end{center}
\caption{Closes paths $C$ and $C^\prime$ in the complex $Q$ plain.
}
\label{fig:path}
\end{figure}
In this appendix, we evaluate the $Q$ integral in the following form
\beqa
I = \int_0^\infty dQ \frac{ J_{L_D}(Q R)}{Q_A^2 -Q^2 + i\varepsilon}  H_{[L],[K]}(Q,Q_{A}),
\eeqa 
for large $R$, assuming that $H_{[L],[K]}(Q,Q_{A})$ have no poles in the  real axis at $Q \ge 0$.
Using the asymptotic form of $ J_{L_D}(x)$ at large $R$ given in eq.~(\ref{eq:bessel_asym}), we write
\beqa
I &\simeq& \sqrt{\frac{2}{\pi R}} \frac{1}{2i} ( I_+-I_-),
\quad R\rightarrow \infty
\eeqa
where
\beqa
I_{\pm} &=&  \int_0^\infty \frac{e^{\pm i (Q R -\Delta_L)}}{Q_A^2-Q^2+i\varepsilon} f(Q), \quad
f(Q) \equiv  \sqrt{\frac{1}{Q}} H_{[L],[K]}(Q,Q_{A}).
\label{eq:Ipm}
\eeqa

We evaluate $I_+$ and $I_-$ separately.
For $I_+$,  we consider
an integration in the complex $Q$ plane on a closed path $C = [0,\infty]  \oplus C_\theta \oplus i[\infty, 0] $ in Fig.~\ref{fig:path}, which leads to
\beqa
I_+ +I_1+I_2 &=& \int_{C}  \frac{e^{ i (Q R -\Delta_L)}}{Q_A^2-Q^2+i\varepsilon} f(Q) = -\frac{\pi i}{Q_A} e^{ i (Q_A R -\Delta_L)} f(Q_A) + O(e^{-cR}),
\eeqa
where 
\beqa
I_1 &\equiv & \lim_{q\rightarrow\infty}\int_0^{\pi/2}  qe^{i\theta}i d\theta\left.
 \frac{e^{ i (Q R -\Delta_L)}}{Q_A^2-Q^2+i\varepsilon} f(Q)\right\vert_{Q=qe^{i\theta}} , \\
I_2 & \equiv & \int_{i\infty}^{i0} dQ  \frac{e^{ i (Q R -\Delta_L)}}{Q_A^2-Q^2+i\varepsilon} f(Q)
= -\int_0^\infty idq \frac{e^{-qR-i\Delta_L}}{Q_A^2+q^2+i\varepsilon}f(iq),
\eeqa
and the term $O(e^{-cR})$ with $c> 0$ represents the contributions from complex poles inside $C$.
It is easy to show that $I_1$ vanishes as
\beqa
\vert I_1\vert &\le & \lim_{q\rightarrow\infty}\underbrace{ \frac{q}{(q^2-Q_A^2)} \max_{0\le \theta\le \pi/2}\left\vert f(qe^{i\theta}) \right\vert}_{\equiv F(q)}  \int_0^{\pi/2} d\theta e^{-qR\sin\theta} \le
\lim_{q\rightarrow\infty} F(q) \int_0^{\pi/2} d\theta e^{-2qR\theta/\pi} \nn \\
&=& \lim_{q\rightarrow\infty} F(q) \frac{\pi}{2qR}( 1-e^{-qR})\rightarrow 0,
\eeqa
where we assume that $\displaystyle \max_{0\le \theta\le \pi/2}\left\vert f(qe^{i\theta}) \right\vert$ does not grow as fast as $q^2$ in the large $q^2$ limit.
Similarly, we estimate
\beqa
\vert I_2 \vert &\le& \int_0^\infty dq\, \frac{e^{-qR}}{Q_A^2+q^2} \vert f(iq) \vert
\le \frac{1}{Q_A^2} \max_{0< q } \vert f(iq)\vert  \int_0^\infty dq\, e^{-qR}
= \frac{1}{Q_A^2} \max_{0< q } \vert f(iq)\vert  \frac{ 1}{R}
\eeqa
for $Q_A\not= 0$, which vanishes $1/R$ for large $R$  as long as $\displaystyle \max_{0< q } \vert f(iq)\vert < \infty$.
If some poles happen to exist on the positive imaginary axis, we can modify  the path a little to avoid poles, so that the above estimate still holds.  
We indeed have a more stronger bound of $\vert I_2\vert$ for all $Q_A$ including $Q_A=0$ as shown below at $n\ge 3$.
(At $n=2$, we can evaluate $I$ by the different method. ) 
Since we can write $f(Q) = Q^{(D-1)/2} g(Q)$ with $\vert g(0) \vert < \infty$ from eqs.~(\ref{eq:defH}) and (\ref{eq:Ipm}), 
we have
\beqa
\vert I_2 \vert &\le&  \max_{0< q } \vert g(iq)\vert \int _0^\infty dq\, q^{(D-5)/2} e^{-qR}
= \max_{0< q } \vert g(iq)\vert R^{(3-D)/2} \int_0^\infty dt\, t^{(D-5)/2}e^{-t} ,
\eeqa
which vanishes as  $R^{(3-D)/2}$ for large $R$ at $ n\ge 3 $ ( $D \ge 6$ ), as long as $\displaystyle \max_{0< q } \vert g(iq)\vert < \infty$. (Again we can modify the path if poles exits on  the positive imaginary axis.)
Altogether we obtain
\beqa
I_+ &\simeq & -\frac{\pi i}{Q_A} e^{i(Q_A R -\Delta_L)} f(Q_A) + O( R^{(3-D)/2} ). 
\eeqa

For $I_-$, we take another closed path $C^\prime =[0,\infty]\oplus C^\prime_\theta \oplus i[-\infty, 0]$ in Fig.~\ref{fig:path}.
Since poles at $Q=\pm(Q_A+i\varepsilon)$ are not contained in this closed path, we have
\beqa
I_+ + I_1^\prime + I_2^\prime &=&  \int_{C^\prime}  \frac{e^{- i (Q R -\Delta_L)}}{Q_A^2-Q^2+i\varepsilon} f(Q) =  O(e^{-c^\prime R})
\eeqa
with $c^\prime > 0$, where
\beqa
I_1^\prime &\equiv & \lim_{q\rightarrow\infty}\int_0^{-\pi/2}  qe^{i\theta}i d\theta\left.
 \frac{e^{- i (Q R -\Delta_L)}}{Q_A^2-Q^2+i\varepsilon} f(Q)\right\vert_{Q=qe^{i\theta}} , \\
I_2^\prime & \equiv & \int_{-i\infty}^{-i0} dQ  \frac{e^{- i (Q R -\Delta_L)}}{Q_A^2-Q^2+i\varepsilon} f(Q)
= \int_0^\infty idq \frac{e^{-qR+i\Delta_L}}{Q_A^2+q^2+i\varepsilon}f(-iq) .
\eeqa
As in the case before, it is easy to show that
\beqa
\vert I_1^\prime \vert  &=& 0, \qquad
\vert I_2^\prime \vert  =   O(R^{(3-D)/2}),
\eeqa
which leads to $I_- = O(R^{(3-D)/2})$.

Combining these, we finally obtain
\beqa
I &=& -\sqrt{\frac{2}{\pi Q_A R}}\left[ \frac{\pi e^{i(Q_A R - \Delta_L)}}{2Q_A} H_{[L],[K]}(Q_A,Q_A) + O
(R^{(3-D)/2})\right ] ,
\eeqa
which proves eq.~(\ref{eq:integral} ).

\section{Coupled channel cases}
\label{sec:coupled}
In this appendix, we extend our investigation to the case where $l\rightarrow n$ scatterings with $l\not= n$
can occur.
\subsection{Unitarity constraint to $T$-matrix}

The unitarity relation to $T$-matrix in eq.~(\ref{eq:unitarity}) can be generalized to 
\beqa
T^\dagger( \bQ_n,  \bQ_l ) &-& T ( \bQ_n,  \bQ_l ) = \sum_k \frac{i}{k^{3/2}}
\int d\bQ_k \, \delta(E_{\bQs_n}-E_{\bQs_k}) \nn \\
&\times& T^\dagger( \bQ_n,  \bQ_k )  T( \bQ_k,  \bQ_l ) 
\label{eq:unitarity_nl}
\eeqa
for general $n, l$, where the energy conservation that $E_{\bQs_n} = E_{\bQs_l}$ is always satisfied. 

As in the case of the single channel, we expand $T$ in term of the hyper-spherical harmonic function as 
\beqa
T ( \bQ_n,  \bQ_l )  &=&  \sum_{[N_n], [L_l]} T_{[N_n], [L_l]}(Q_n, Q_l) Y_{[N_n]}(\Omega_{\bQs_n})
\overline{Y_{[L_l]}(\Omega_{\bQs_l})},
\eeqa
where $ Q_n^2- Q_l^2 = 2m^2 (l-n) $ in the non-relativistic approximation.
Putting this into eq.~(\ref{eq:unitarity_nl}), we have
\beqa
T_{[N_n], [L_l]}^\dagger(Q_n, Q_l)-T_{[N_n], [L_l]}(Q_n, Q_l)  &=&\sum_{k, [K_k]}
i\frac{ m Q_k^{D_k-2}}{k^{3/2}}  T_{[N_n], [K_k]}^\dagger(Q_n, Q_k) T_{[K_k], [L_l]}(Q_k, Q_l) \nn \\
\label{eq:unitarity_nl_B}
\eeqa
where $D_k = 3(k-1)$ and  $ Q_n^2- Q_k^2 = 2m^2 (k-n) $.  Defining and diagonalizing $\hat T$ as
\beqa
\hat T_{[N_n], [L_l]}(Q_n, Q_l) &\equiv &  \frac{Q_n^{D_n/2-1}}{n^{3/4}}T_{[N_n], [L_l]}(Q_n, Q_l) \frac{Q_l^{D_{l}/2-1}}{l^{3/4}} \nn \\
&=& \sum_{k, [K_k]} U_{[N_n], [K_k]}(Q_k) \hat T_{[K_k]}(Q_k) U_{[K_{k}], [L_l]}^\dagger(Q_k) ,
\eeqa 
where $ Q_n^2- Q_k^2 = 2m^2 (k-n) $, eq.~(\ref{eq:unitarity_nl_B}) leads to
\beqa
 \hat T_{[K_k]}(Q_k) &=& - \frac{2}{m}e^{i\delta_{[K_k]}(Q_k)}\sin\delta_{[K_k]}(Q_k) .
\eeqa
This gives us the final result,
\beqa
T_{[N_n], [L_l]}(Q_n, Q_l) &=&- \frac{2 n^{3/4} l^{3/4}}{m Q_n^{D_n/2-1} Q_l^{D_l/2-1}}
 \sum_{k, [K_k]} U_{[N_n], [K_k]}(Q_k) e^{i\delta_{[K_k]}(Q_k)}\sin\delta_{[K_k]}(Q_k) \nn \\
 &\times& U_{[K_{k}], [L_l]}^\dagger(Q_k) ,
\eeqa
which reproduces eq.~(\ref{eq:Tn}) for the single channel at $n=l=k$.

\subsection{Asymptotic behavior of the NBS wave function}
For the coupled channel, the NBS wave function corresponding to eq.~(\ref{eq:NBS_nonR}) in the non-relativistic approximation becomes 
\beqa
\Psi^{nl}(\bR_n, \bQ_l) &=& C_n \left[ \delta_{nl}\, e^{i\bQs_l\cdot\bRs_n} +\frac{2m}{2\pi n^{3/2}}
\int d\bP_n \frac{e^{i\bPs_n\cdot\bRs_n} T(\bP_n,\bQ_l)}{Q_l^2-P_n^2+2m^2(l-n)+i\varepsilon}
\right]  ,
\eeqa
where $C_n = ((2\pi)^3  2m )^{-n/2}$. (We here omit irrelevant $\dfrac{Q_n^2}{m}$ contributions.)
Expanding the NBS wave function in terms of the hyper-spherical function as
\beqa
\Psi^{nl}(\bR_n, \bQ_l) &=& \sum_{[N_n],[L_l]} \Psi_{[N_n],[L_l]} (R_n, Q_l) Y_{[N_n]}(\Omega_{\bRs_n})
\overline{Y_{[L_l]}(\Omega_{\bQs_l})},
\eeqa 
together with Eq.~(\ref{eq:plain_D}), we have
\beqa
\Psi_{[N_n],[L_l]} (R_n, Q_l) &=& C_n i^{N_n} \frac{(2\pi)^{D_n/2}}{ (Q_n R_n)^{D_n/2-1}}
\left[ J_{\tilde N_n}(Q_n R_n)\delta_{nl}\delta_{[N_n],[L_l]}\right. \nn \\ 
&+& \left. \int dP_n\, \frac{J_{\tilde N_n}(P_n R_n)}{Q_l^2-P_n^2+2m^2(l-n)+i\varepsilon}
H_{[N_n],[L_l]}(P_n, Q_l)
\right]
\label{eq:NBS_couple0}
\eeqa
where $\tilde N_n = N_n + (3n-5)/2$ and 
\beqa
H_{[N_n],[L_l]}(P_n, Q_l) &=& \frac{m}{\pi n^{3/2}} P_n^{D_n/2} Q_n^{D_n/2-1}
T_{[N_n],[L_l]}(P_n, Q_l) .
\eeqa
As before, after $P_n$ integral, the second term in eq.~(\ref{eq:NBS_couple0}) for large $R_n$ is given by
\beqa
&\simeq&\left[ H_{\tilde N_n}(Q_n R_n) + i J _{\tilde N_n}(Q_n R_n) \right]  \left(\frac{l}{n}\right)^{3/4}
\frac{Q_n^{D_n/2-1}}{Q_l^{D_l/2-1}}\nn \\
&\times&  \sum_{k, [K_k]}
U_{[N_n],[K_k]}( Q_k) e^{i\delta_{[K_k]}(Q_k)} \sin \delta_{[K_k]}(Q_k) U_{[K_k],[L_l]}^\dagger(Q_k),
\eeqa
where $Q_n^2 = Q_l^2 + 2m^2(l-n)$ and $Q_k^2 = Q_l^2 + 2m^2(l-k)$. We therefore finally obtain
\beqa
\Psi_{[N_n],[L_l]} (R_n, Q_l) &\simeq& C_n i^{N_n} \frac{(2\pi)^{D_n/2}}{ (Q_n R_n)^{D_n/2-1}} \left(\frac{l}{n}\right)^{3/4}
\frac{Q_n^{D_n/2-1}}{Q_l^{D_l/2-1}}
 \sum_{k, [K_k]} U_{[N_n],[K_k]}(Q_k) e^{i\delta_{[K_k]}(Q_k)}
   \nn \\
&\times &\left[  J_{\tilde N_n}(Q_n R_n) \cos \delta_{[K_k]}(Q_k) + H_{\tilde N_n}(Q_n R_n)\sin \delta_{[K_k]}(Q_k)  \right] U_{[K_k],[L_l]}^\dagger(Q_k) \nn \\
&\simeq & 
C_n i^{N_n} \frac{(2\pi)^{D_n/2}}{ (Q_n R_n)^{D_n/2}} \left(\frac{l}{n}\right)^{3/4}
\frac{Q_n^{D_n/2-1}}{Q_l^{D_l/2-1}}
 \sum_{k, [K_k]} U_{[N_n],[K_k]}(Q_k) e^{i\delta_{[K_k]}(Q_k)} \nn \\
&\times &  \sqrt{\frac{2}{\pi}}\sin (Q_n R_n -\Delta_{N_n} +\delta_{[K_k]}(Q_k))\  U_{[K_k],[L_l]}^\dagger(Q_k) 
\eeqa
where $\Delta_{N_n} = (2 \tilde N_n-1)\pi/4$, 
which correctly reproduces eq.~(\ref{eq:main_result}) in the single channel case at $n=l=k$.

\end{document}